\documentclass{article}%
\usepackage{amsfonts}
\usepackage{amsmath}
\usepackage{amssymb}
\usepackage{graphicx}%
\begin{document}

\title{Conformally flat spacetimes and Weyl frames }
\author{C. Romero, J. B. Fonseca-Neto and M. L. Pucheu\\$^{a}$Departamento de F\'{\i}sica, Universidade Federal da Para\'{\i}ba,\\C.Postal 5008, 58051-970 Jo\~{a}o Pessoa, Pb, Brazil\\E-mail: cromero@fisica.ufpb.br}
\maketitle

\begin{abstract}
We discuss the concepts of Weyl and Riemann frames in\ the context of metric
theories of gravity and state the fact that they are completely equivalent as
far as geodesic motion is concerned. We apply this result to conformally flat
spacetimes and show that a new picture arises when a Riemannian spacetime is
taken by means of geometrical gauge transformations into a Minkowskian flat
spacetime. We find out that in the Weyl frame gravity is described by a scalar
field. We give some examples of how conformally flat spacetime configurations
look when viewed from the standpoint of a Weyl frame. \ We show that in the
non-relativistic and weak field regime the Weyl scalar field may be identified
with the Newtonian gravitational potential. We suggest an equation for the
scalar field by varying the Einstein-Hilbert action restricted to the class of
conformally-flat spacetimes. We revisit Einstein and Fokker's interpretation
of Nordstr\"{o}m scalar gravity theory and draw an analogy between this
approach and the Weyl gauge formalism. We briefly take a look at
two-dimensional gravity\ as viewed in the Weyl frame and address the question
of quantizing a conformally flat spacetime by going to the Weyl frame.

\end{abstract}

\section{Introduction}

It is well known that the\ concept of \textit{geodesics }plays a role of
fundamental importance in general relativity as well in any metric theory of
gravity. Indeed, an elegant aspect of the geometrization of the gravitational
field lies in the geodesics postulate, that is, the statement that light
rays\ and particles moving under the influence of gravity alone follow
spacetime geodesics. This fact means that a great deal of information about
the motion of particles in a given spacetime is promptly available once one
knows its geodesic structure, i.e the set of all geodesics admitted by that
spacetime. In general relativity, geodesics are completely determined by the
metric properties of the spacetime since it is also assumed that the spacetime
geometry has a Riemannian character \cite{Einstein}. However, in\ many other
metric theories of gravity one distinguishes between \textit{metric geodesics
}and \textit{affine geodesics}, and so\textit{ }in these theories one must be
careful, from the outset, to clearly specify to which kind of geodesics does
the geodesic postulates refer \cite{Will}.\textit{ }In any case, it is by
analyzing the behaviour of timelike and null geodesics, the later determining
the light-cone structure, that one is able to predict a series of relativistic
phenomena, such as, the existence of the perihelium precession of Mercury%
\'{}%
s orbit, the deflection of the light by the Sun, the gravitation redshift of
light and the gravitational time delay (Shapiro effect). In addition to these,
almost all the physics of black holes is obtained by studying the properties
of geodesics near the spacetime event horizon. Finally, to explain the
existence of a cosmological redshift, the acceleration of the Universe, and
many other empirical facts of cosmology all we need is to know the
mathematical behaviour of the geodesics corresponding to the underlying
cosmological model.

In view of the above, we may conclude that as far as the information conveyed
by the geodesic lines of a certain spacetime is concerned one has a certain
degree of freedom in the choice of the geometry associated with that space.
For instance, in a certain sense it does not seem that the concept of
\textit{Riemannian}\ \textit{curvature} is essential for the geometrical
description of the gravitational and cosmological phenomena just mentioned.
Two distinct geometries sharing the same geodesic structure will give exactly
the same description of geodesic-related phenomena, being for this reason
indistinguishable from the observational point of view. In this sense the two
geometries may be regarded as equivalent. If, in addition, they are related by
some kind of mathematical transformation, it may happen that one of them is
preferable to the other when we need to do some calculations, or if we want to
get a simpler or different picture of physical processes going on. \ In this
paper, we would like to develop further these ideas by considering a kind of
interplay between two different frameworks: the geometries of Riemann and
Weyl. As we will see, there are circumstances in which it is possible to swift
from one to the other while keeping some basic geometric structure invariant.
The key notion to understand how such correspondence works is that of
geometrical \textit{gauge transformation}, a concept introduced by H. Weyl in
1918 \cite{Weyl}. The theory developed by Weyl is regarded by many as an
elegant generalization of Riemannian geometry, and, in the opinion of some
authors, \ "contains a suggestive formalism and may still have the germs of a
future fruitful theory " \cite{Bazin}.

This paper is organized as follows. In Section II, we give a brief
introduction to Weyl geometry and introduce the notion of Riemann and Weyl
frames. We proceed in Section III to consider how the class of conformally
spacetimes is described in the Weyl frame. In Section IV, we show that in the
weak field regime the scalar field that appears in the Weyl frame may be
identified with the Newtonian gravitational potential. Then, in Section V, we
suggest an equation for the Weyl scalar field which may be deduced by varying
the Einstein-Hilbert action with respect to the restricted class of
conformally flat metrics.\ Section VI contains a brief discussion of the
analogy between the dynamics of the Weyl scalar field and the approach
followed by\ Nordstr\"{o}m scalar theory of gravity. In Sections VII we take a
brief look at two-dimensional gravity\ as viewed in the Weyl frame. Finally,
in Section VII, we address the question of quantizing a conformally flat
spacetime by going to the Weyl frame where the problem may be reduced to the
quantization of a scalar field in flat spacetime. We summarize our work in
Section IX.

\section{Weyl geometry}

Conceived by Weyl in 1918, as an attempt to unify gravity with
electromagnetism, in its original form Weyl%
\'{}%
s theory \cite{Weyl} turned out to be inadequate as a physical theory as was
firstly pointed by Einstein soon after the appearance of the theory
\cite{Pauli}.\ As is well known, Einstein's argument was that in a
non-integrable\ Weyl geometry it would not be possible the existence of sharp
spectral lines in the presence of an electromagnetic field since atomic clocks
would depend on their past history, an effect known as \textit{the second
clock effect}.\ However, a variant of Weyl geometry,\ namely, the one in which
the Weyl field is integrable, does not suffer from this flaw \cite{Perlick},
and, for this reason, has been used in some approaches to gravitation and
cosmology in varied contexts \cite{Novello}. Interest in Weyl geometry among
physicists has also been increased by the constructive-axiomatic formulation
of spacetime theory developed by Ehlers, Pirani and Schild, who
demonstrated\ that if certain axioms, suggested by experience, are satisfied,
then one is naturally led to Weyl geometry \cite{Ehlers}. On the other hand,
there are arguments based on quantum mechanics that seems to rule out
non-integral Weyl geometry as a viable \cite{Ausdretsch} framework to describe
spacetime, although this point remains controversial \cite{Perlick}.

Let us now discuss what kind of geometry Weyl discovered. The essential
difference between the Riemann geometry and the Weyl geometry is that in the
former one makes the assumption that the covariant derivative $\nabla
_{a}g_{bc}$ of the metric tensor $g$ is zero, while in the latter $\nabla
_{a}g_{bc}$ is given by%

\begin{equation}
\nabla_{a}g_{bc}=\sigma_{a}g_{bc} \label{compatibility}%
\end{equation}
where $\sigma_{a}$ \ denotes the components of a one-form field $\sigma\ $with
respect to a local coordinate basis. This represents a generalization of the
Riemannian condition of compatibility between the connection $\nabla$ and $g,$
which is equivalent to require the length of a vector to remain unaltered by
parallel transport \cite{Pauli}. If $\sigma$ is an exact form, i.e. $\sigma$
$=d\phi,$ where $\phi$ is a scalar field, then we have what is called\ an
integrable Weyl geometry. The triad $(M,g,\sigma)$ where $M$ is a
differentiable manifold endowed with a metric $g$ and a Weyl field $\sigma
\ $will be referred to as a \textit{Weyl frame}. It is interesting to note
that the Weyl condition (\ref{compatibility}) remains unchanged\ when we go to
another Weyl frame $(M,\overline{g},\overline{\sigma})$ by performing the
following simultaneous transformations in $g$ and $\sigma$:%
\begin{equation}
\overline{g}=e^{-f}g \label{conformal}%
\end{equation}%
\begin{equation}
\overline{\sigma}=\sigma-df \label{gauge}%
\end{equation}
where $f$ is a scalar function defined on $M$.

Quite analogously to Riemannian geometry, the condition (\ref{compatibility})
is sufficient to completely determine\ the Weyl connection\ $\nabla$\ in terms
of the metric $g$\ and the Weyl one-form field $\sigma.$ Indeed, a
straightforward calculation shows that one can express the components of the
affine connection with respect to an arbitrary vector basis completely in
terms of the components of $g$ and $\sigma$:%
\begin{equation}
\Gamma_{\beta\gamma}^{\alpha}=\{_{\beta\gamma}^{\alpha}\}-\frac{1}{2}%
g^{\alpha\mu}[g_{\mu\beta}\sigma_{\gamma}+g_{\mu\gamma}\sigma_{\beta}%
-g_{\beta\gamma}\sigma_{\mu}] \label{Weylconnection}%
\end{equation}
where $\{_{bc}^{a}\}$ represents the Christoffel symbols.

A clear geometrical insight on the properties of Weyl parallel transport is
given by the following proposition: Let $M$ be a differentiable manifold with
an affine connection $\nabla$, a metric $g$ and a Weyl field of one-forms
$\sigma$. If $\nabla$ is compatible with $g$ in the Weyl sense,\ i.e. if
(\ref{compatibility}) holds, then for any smooth curve $\alpha=\alpha
(\lambda)$ and any pair of two parallel vector fields $V$ and $U$ along
$\alpha,$ we have
\begin{equation}
\frac{d}{d\lambda}g(V,U)=\sigma(\frac{d}{d\lambda})g(V,U)
\label{covariantderivative}%
\end{equation}
where $\frac{d}{d\lambda}$ denotes the vector tangent to $\alpha$.

If we integrate the above equation along the curve $\alpha$, starting from a
point $P_{0}=\alpha(\lambda_{0}),$ then we obtain%
\begin{equation}
g(V(\lambda),U(\lambda))=g(V(\lambda_{0}),U(\lambda_{0}))e^{\int_{\lambda_{0}%
}^{\lambda}\sigma(\frac{d}{d\rho})d\rho} \label{integral}%
\end{equation}
Putting $U=V$ and denoting by $L(\lambda)$ the length of the vector
$V(\lambda)$ at an arbitrary point\ $P=\alpha(\lambda)$\ of the curve, then it
is easy to see that in a local coordinate system $\left\{  x^{a}\right\}  $
the equation (\ref{covariantderivative}) reduces to
\[
\frac{dL}{d\lambda}=\frac{\sigma_{\alpha}}{2}\frac{dx^{\alpha}}{d\lambda}L
\]

Consider the set of all closed curves $\alpha:[a,b]\in R\rightarrow M$, i.e,
with $\alpha(a)=\alpha(b).$ Then, \ we have the equation
\[
g(V(b),U(b))=g(V(a),U(a))e^{\int_{a}^{b}\sigma(\frac{d}{d\lambda})d\lambda}.
\]
Now, it is the integral\ $^{\int_{a}^{b}\sigma(\frac{d}{d\lambda})d\lambda}$
that is responsible for the difference between the readings of two identical
atomic clocks following different paths. It follows from Stokes' theorem that
if $\sigma$ is an exact form, that is, \ if there exists a scalar function
$\phi$, such that $\sigma=d\phi$, then%

\begin{equation}
\oint\sigma(\frac{d}{d\lambda})d\lambda=0 \label{int}%
\end{equation}
for any loop. In other words, in this case the integral $\int_{a}^{b}%
\sigma(\frac{d}{d\rho})d\rho$\ does not depend on the path.\ Since it is this
integral that regulates the way atomic clocks run this variant of Weyl
geometry does not suffer from the flaw pointed out by Einstein, and we have
what is often called in the literature a \textit{Weyl integrable manifold}.

Another way to look at (\ref{int}) is the following. From Frobenius' theorem
we know that $\sigma$ is an exact form if and only if $d\sigma=0$. In local
coordinates where $\sigma=\sigma_{\alpha}dx^{\alpha}$ this condition reads
$F_{\alpha\beta}=$ $\sigma_{\alpha,\beta}-$ $\sigma_{\beta,\alpha}=0$. The
quantity $F_{\alpha\beta}=\sigma_{\alpha,\beta}-$ $\sigma_{\beta,\alpha}$ ,
which is non-vanishing in general, is easily shown to be gauge invariant and
was interpreted by Weyl as the electromagnetic field in his attempt to
geometrize electromagnetism \cite{Weyl}. The $2$-form $F=F_{\alpha\beta
}dx^{\alpha}\symbol{94}dx^{\beta}$ is called \textit{length curvature}, so a
Weyl integrable manifold is one in which the length curvature $F$ vanishes.
Finally, if there is a frame $(M,g,\sigma)$ in which $\sigma=0$, then
obviously the geometry of Weyl reduces to Riemann geometry\ in that frame.

An important feature of Weyl geometry, which will be explored in this paper,
is the following mathematical fact. Consider the (affine) geodesic equations
\begin{equation}
\nabla_{V}V=0 \label{geodesic}%
\end{equation}
in a certain frame $(M,g,\sigma)$, where $V$ denotes the tangent vector to the
geodesic curve. In local coordinates (\ref{geodesic}) has the form
\begin{equation}
\frac{d^{2}x^{\mu}}{d\lambda^{2}}+\Gamma_{\;\alpha\beta}^{\mu}\frac
{dx^{\alpha}}{d\lambda}\frac{dx^{\beta}}{d\lambda}=0 \label{geodesics1}%
\end{equation}
where $\Gamma_{\;\alpha\beta}^{\mu}$ denotes the components of the connection
$\nabla$, $\lambda$ is an affine parameter and $x^{\mu}=x^{\mu}(\lambda)$
represents local parametric equations of the geodesic. Suppose that we change
from the frame $(M,g,\sigma)$ to another frame $(M,\overline{g},\overline
{\sigma})$ by performing \ a gauge transformation in accordance with
(\ref{conformal}) and (\ref{gauge}). It is clear that in each frame the
components of $\nabla$, i.e $\Gamma_{\;\alpha\beta}^{\mu}$ and $\overline
{\Gamma}_{\;\alpha\beta}^{\mu}$, may be expressed in terms of the Christoffel
symbols $\{_{\beta\gamma}^{\alpha}\}$, $\overline{\{_{\beta\gamma}^{\alpha}%
\}}$ \ and the Weyl fields $\sigma$ and $\overline{\sigma}$, respectively, as
in Eq. (\ref{Weylconnection}) . Because $\nabla$ is kept unaltered by the
gauge transformations, if $x^{\mu}=x^{\mu}(\lambda)$ is a solution of
(\ref{geodesic}) in the frame $(M,g,\sigma)$, then it is also a solution of
that equation in the other frame $(M,\overline{g},\overline{\phi})$. The
geodesic equations are gauge invariant because $\Gamma_{\;\alpha\beta}^{\mu}$
$=$ $\overline{\Gamma}_{\;\alpha\beta}^{\mu}$, and the truth of this statement
can be \ easily verified by explicitly using (\ref{Weylconnection}).

\section{Conformally flat spacetimes and the Weyl gauge field}

In the light of the concepts just discussed let us consider in this section
the class of all conformally flat spacetimes. As we know, a significant number
of spacetimes of physical interest predicted by general relativity belong to
this class. For instance, it is well known that all Robertson-Walker
cosmological models are conformally flat. Explicit conformal transformations
taking these to flat spacetime were first given by Infeld and Schild
\cite{Infeld}. Let us consider more generally a certain conformally flat
spacetime $M$ with a metric $g=e^{\phi}\eta$, where $\eta$ denotes the
Minkowski metric and $\phi$ is a scalar function. If the geometry of $M$ is
Riemannian we have no Weyl field, and\ so the components of the affine
connection $\Gamma_{\;\alpha\beta}^{\mu}$ are identical to the Christoffel
symbols $\{_{\alpha\beta}^{\mu}\}$. On the other hand,\ this geometrical
configuration is clearly equivalent to the one described in terms of Weyl
geometry as long as we confine ourselves to the frame $(M,e^{\phi}\eta,0)$, to
which we will refer as the \textit{Riemann frame.} Suppose now that we make
the gauge transformation (\ref{conformal}) and (\ref{gauge}) with $f$
\ replacing $\phi$. In doing so, we arrive at a frame, namely $(M,\eta
,-d\phi)$, which will be called the \textit{Weyl frame.} As we have seen, with
respect to geodesics both frames are completely equivalent. Nevertheless, in
many aspects the geometry defined by them are entirely distinct. For instance,
from the point of view of the Riemannian\ curvature they are obvioulsly
distinct. In effect, in the Riemann frame $(M,e^{\phi}\eta,0)$ the manifold
$M$ is endowed with a metric that leads to non-zero curvature, while in the
Weyl frame $(M,\eta,-d\phi)$ we\ have a flat spacetime (in the Riemannian
sense). Another difference concerns the length of non-null curves or other
metric-dependent geometrical quantities since in the two frames we have
distinct metric tensors. Null curves, on the other hand, are mapped into null
curves. This implies that the light geometry of a conformally flat spacetime
is identical to that of Minkowski spacetime, a well known feature of conformal
transformations. It is here that the geometrical framework conceived by Weyl
comes into play.

As we have already remarked, from the standpoint of geodesics and the
lightcone structure we can characterize any cosmological model whose geometry
is conformal purely in terms of an integrable Weyl field $\sigma=d\phi$
defined in Minkowski spacetime. It turns out that the passage from the frame
$(M,e^{\phi}\eta,0)$ to the frame $(M,\eta,-d\phi)$ provides us with a new
geometrical picture. For instance, cosmological phenomena, such as\ the
redshift of galaxies or the expansion of the Universe, cease to be necessarily
explained by the action of a dynamical curved spacetime . Instead, the
dynamics of the Cosmos becomes, in this picture, entirely governed by a gauge
field living in a fixed and static spacetime. It is this gauge field, by the
way, that controls a new law of parallel displacement and also determines the
behaviour of clocks and measuring devices. In other words, to each conformal
Riemannian spacetime we can associate a Weyl gauge field in Minkowski
spacetime, whose dynamics is ultimately the sole responsible for the motion of
particles and light rays. Let us illustrate this point by explicitly
calculating the Weyl field from some simple cosmological models.

Consider a Robertson-Walker metric $g$ corresponding to a homogeneous and
isotropical cosmological model, written in the form\footnote{Throughout this
paper, except in Section IV, we set $c=1$. We are also adopting the following
convention in the definition of the Riemann and Ricci tensors: $R_{\;\mu
\beta\nu}^{\alpha}=\Gamma_{\beta\mu,\nu}^{\alpha}-\Gamma_{\mu\nu,\beta
}^{\alpha}+\Gamma_{\rho\nu}^{\alpha}\Gamma_{\beta\mu}^{\rho}-\Gamma_{\rho
\beta}^{\alpha}\Gamma_{\nu\mu}^{\rho};$ $R_{\mu\nu}=R_{\;\mu\alpha\nu}%
^{\alpha}.$ In this convention, we will write the Einstein equations as
$R_{\mu\nu}-\frac{1}{2}Rg_{\mu\nu}-\Lambda g_{\mu\nu}=-\kappa T_{\mu\nu},$
with $\kappa=\frac{8\pi G}{c4}$.}
\begin{equation}
ds^{2}=dt^{2}-A^{2}(t)(\frac{dr^{2}}{1-kr^{2}}+r^{2}d\theta^{2}+r^{2}\sin
^{2}\theta d\varphi^{2})\text{ ,}\label{FRW}%
\end{equation}
where $k=0,\pm1.$ Taking, for simplicity, the case of flat spatial section
$(k=0)$, and defining the \ so-called \textit{conformal time} $d\tau=\frac
{dt}{A(t)}$, we can rewrite \ the metric (\ref{FRW}) in the \ conformally flat
form
\begin{equation}
ds^{2}=S^{2}(\tau)(d\tau^{2}-dr^{2}-r^{2}d\theta^{2}-r^{2}\sin^{2}\theta
d\varphi^{2})\text{ ,}\label{FRW1}%
\end{equation}
where we have defined \ $S(\tau)=A(t(\tau))$. By carrying out the
transformations (\ref{conformal}) and (\ref{gauge}) $\ $with $e^{f}=S^{2}%
(\tau)$, we can pass from the Riemann frame $(M,g,0)$ to the Weyl frame
$(M,\eta,\sigma)=\ (M,\eta,$ $-d(2\ln S(\tau))$. In this frame, the Weyl gauge
field is given by the $1$-form $\sigma=\frac{-2}{S}\frac{dS}{d\tau}d\tau$,
whose components in the coordinate basis are $\sigma_{\mu}=($ $\frac{-2}%
{S}\frac{dS}{d\tau},0,0,0).$ In the case of spatially flat Friedmann models,
we have $A(t)=A_{0}$\bigskip$t^{p}$ , where $A_{0}$ is a constant. This
functional form of $A(t)$ includes the so-called matter-dominated $(p=\frac
{2}{3})$ and the radiation-dominated $(p=\frac{1}{2})$ universes, and a
possible choice for the conformal time is $\tau=\frac{1}{A_{0}(1-p)}t^{1-p}$ (
$p\neq1$). The expression for the Weyl scalar field $\phi$ as a function of
$\tau$ then becomes
\[
\phi(\tau)=-2\ln A(t(\tau))=-2\ln(a_{0}\tau)^{\frac{p}{1-p}}\text{ ,}%
\]
where we have defined $a_{0}=(1-p)(A_{0})^{\frac{1}{p}}$. \ Thus, for the
matter-dominated and the radiation-dominated universes, the Weyl scalar field
$\phi(\tau)$\ will be given, respectively, by $\phi(\tau)=-4\ln a_{0}\tau$ and
$\phi(\tau)=-2\ln a_{0}\tau$. We note that in both cases $\phi(\tau)$ has a
singularity at $\tau=0$.

Another simple example is given by the de Sitter-Lema\^{\i}tre cosmological
model, whose metric is given by (\ref{FRW}), with $k=0$ and $A(t)=A_{0}%
e^{\sqrt{\frac{\Lambda}{3}}t}$, where $\Lambda$ is a positive constant. If we
choose the conformal time as $\tau=-\frac{1}{A_{0}}\sqrt{\frac{3}{\Lambda}%
}e^{-\sqrt{\frac{\Lambda}{3}}t}$, then the scalar \ field will be given by
\[
\phi(\tau)=-2\ln(-\sqrt{\frac{\Lambda}{3}}\frac{1}{\tau}).
\]
In local coordinates, we have $\sigma_{\mu}=(\frac{2}{\tau},0,0,0)$.

It has been shown recently that the metric of all Robertson-Walker\ (RW)
models ($k=0,\pm1$) is conformally flat \cite{Ibison}. For each of these we
may apply the above procedure to\ obtain the Weyl scalar field $\phi$, hence
the gauge field $\sigma=d\phi$ , and for $k=\pm1$ both will be a function of
$r$ and $t$.

In view of the above, we see that in the Weyl frame's picture the kinematical
behaviour of galaxies in any Robertson-Walker cosmological model is totally
determined by the Weyl scalar field $\phi$ while spacetime remains fixed. On
the other hand, $\phi$\ , now looked upon primarily as the conformal factor of
a conformally flat spacetime in the Riemann frame, may be considered as a
gauging function determining the behaviour of clocks and mesuring rods in a
Minkowski spacetime. This second view was noted long ago by Infeld and Schild
(\cite{Infeld}). A similar scenario was conceived more recently in which the
role of the scalar field is replaced by space and time variation of particle
masses. In any of these pictures, the interesting fact is that it is possible
to conceive a new scenario, in which the Riemannian curvature ceases to
determine the cosmic expansion and other cosmological phenomena, which in our
case is the sole responsibility of a scalar field $\phi$.

\section{Gravity in the Weyl frame}

As we have seen, when we go from one frame\ $(M,g,\sigma)$\ to another frame
$(M,\overline{g},\overline{\sigma})$ through the gauge transformations
(\ref{conformal}) and (\ref{gauge}), the pattern of affine\ geodesic curves
does not change. In particular, the metric geodesics\ corresponding to a
conformally flat spacetime in the Riemann frame $(M,e^{\phi}\eta,0)$\ are
completely indistinguishable from the affine geodesics in the Weyl frame
$(M,\eta,-d\phi)$, although in the latter \ a quite different geometrical
picture arises as these curves now lie in a fixed and flat spacetime. This
change of perspective might lead, in some cases, to new insights in the
description of gravitational phenomena. In the case of a conformally flat
spacetime going to the Weyl frame leads to a scenario in which the
gravitational field is not associated with a tensor, but with a geometrical
scalar field $\phi$ living in a Minkowski background. We can get some insight
on the amount of \ physical information carried by the scalar field $\phi$, in
the Weyl frame,\ by investigating the behaviour of $\phi$,in the regime of
weak gravity in the Riemannian frame, where $\phi$ plays the role of a
conformal factor. This is the question we want to examine in this section.

Let us recall that a metric theory of gravity is said to possess a Newtonian
limit in the non-relativistic weak-field regime if one can derive the
Newtonian second law from the geodesic equations as well as the Poisson's
equation from the gravitational field equations. Since in Newtonian physics
the space geometry is Euclidean, a weak gravitational field in a geometric
theory of gravity should manifest itself as a metric phenomenon\ through a
slight perturbation of the Minkowskian spacetime metric. Thus we consider a
time-independent metric tensor of the form
\begin{equation}
g_{\mu\nu}\simeq\eta_{\mu\nu}+\epsilon h_{\mu\nu}\text{ ,}
\label{quasi-Minkowskian}%
\end{equation}
\ where $n_{\mu\nu}$ is the Minkowski tensor, $\epsilon$ is a small parameter
and the term $\epsilon h_{\mu\nu}$ represents a very small time-independent
perturbation due to the presence of some matter configuration. For a
conformally flat spacetime we have $g_{\mu\nu}=e^{\phi}\eta_{\mu\nu}%
\simeq(1+\phi)\eta_{\mu\nu}$. If we adopt the Galilean coordinates of special
relativity we can write the line element defined by (\ref{quasi-Minkowskian})
as
\begin{equation}
ds^{2}=\left(  1+\phi\right)  [(dx^{0})^{2}-(dx^{1})^{2}-(dx^{2})^{2}%
-(dx^{3})^{2}], \label{weakfield}%
\end{equation}
where, as usual, $x^{0}=ct$. Let us now consider the motion of a test particle
in the spacetime (\ref{weakfield}). Since we are working in the
non-relativistic regime we will suppose that the velocity $V^{\alpha}%
=\frac{dx^{\alpha}}{dt}$ of the particle along the geodesic is much less then
$c$, so that the $\beta^{\alpha}=\frac{V^{\alpha}}{c}$ will be regarded as
very small; so in our calculations only first-order terms in $\epsilon$ and
$\beta$ will be kept. Note that in this approximation $\phi$ is regarded as
being static and small, i.e. of the same order as $\epsilon$.

Let us now consider the geodesic equations (\ref{geodesics1})%
\begin{equation}
\frac{d^{2}x^{\mu}}{ds^{2}}+\Gamma_{\;\alpha\beta}^{\mu}\frac{dx^{\alpha}}%
{ds}\frac{dx^{\beta}}{ds}=0, \label{geodesics2}%
\end{equation}
in the Riemann frame $(M,e^{\phi}\eta,0)$, with $\Gamma_{\;\alpha\beta}^{\mu}$
given by (\ref{Weylconnection}).$\ $Because $\Gamma_{\;\alpha\beta}^{\mu}$ is
invariant with respect to the gauge transformations (\ref{conformal}) and
(\ref{gauge})\ the above equations look exactly the same in the Weyl frame
$(M,\eta,-d\phi)$. On the other hand, in view of the fact that in this frame
$\{_{bc}^{a}\}=0$ we have, to first order in $\phi$,
\begin{equation}
\Gamma_{\;\mu\nu}^{\alpha}=\frac{1}{2}n^{\alpha\lambda}[n_{\lambda\mu}%
\phi_{,\nu}+n_{\lambda\nu}\phi_{,\mu}-n_{\mu\nu}\phi_{,\lambda}].
\end{equation}
Recalling that in this approximation
\begin{equation}
\left(  \frac{ds}{dt}\right)  ^{2}\cong c^{2}(1+\epsilon h_{00})=c^{2}%
(1+\phi), \label{propertime}%
\end{equation}
it is not difficult to see that, unless $\mu=\nu=0$, the product
$\Gamma_{\;\alpha\beta}^{\mu}\frac{dx^{\alpha}}{ds}\frac{dx^{\beta}}{ds}$ is
of order $\beta\phi$ or higher. In this way, the geodesic equations
(\ref{geodesics2}) become, to first order in $\phi$ and $\beta$%
\[
\frac{d^{2}x^{\mu}}{ds^{2}}+\Gamma_{\;00}^{\mu}\left(  \frac{dx^{0}}%
{ds}\right)  ^{2}=0
\]
By taking into account (\ref{propertime})\ again, the above equation may be
written as
\begin{equation}
\frac{d^{2}x^{\mu}}{dt^{2}}+c^{2}\Gamma_{\;00}^{\mu}=0
\label{equation-of-motion}%
\end{equation}
Clearly, for $\mu=0$ the equation (\ref{equation-of-motion}) reduces to an
identity. On the other hand, if $\mu$ is a spatial index a simple calculation
gives $\Gamma_{\;00}^{i}=-\frac{\eta^{ij}}{2}\frac{\partial\phi}{\partial
x^{j}}$, and the geodesic equation in this approximation becomes, in
three-dimensional vector notation
\begin{equation}
\frac{d^{2}\overrightarrow{X}}{dt^{2}}=-\frac{c^{2}}{2}\overrightarrow{\nabla
}\phi, \label{NPotential-Rframe}%
\end{equation}
which is simply Newton's equation of motion in a classical gravitational field
provided we identify the scalar gravitational potential as
\begin{equation}
U=\frac{c^{2}}{2}\phi\text{.} \label{Newtonian-potential}%
\end{equation}
Therefore, as regards to the equation of motion of a test particle, we see
that, because Eq. \ref{NPotential-Rframe} also holds in the Weyl frame, the
scalar field $\phi$, when viewed in this frame, plays the role of the
Newtonian gravitational potential.

Our next step is to obtain, still in the weak field approximation, a field
equation for $\phi$. In order to do that we start with the Einstein equations
\begin{equation}
R_{\mu\nu}-\frac{1}{2}g_{\mu\nu}R=-\kappa T_{\mu\nu},
\label{Einstein equation}%
\end{equation}
and find the expressions for $R_{\mu\nu}$ and $R$ when we take $g_{\mu\nu
}=e^{\phi}\eta_{\mu\nu}$. In this way we get \footnote{In $N$ dimensions,\ for
a conformally flat metric $g=e^{\phi}\eta$ the Ricci tensor $R_{\alpha\beta}$
and the scalar curvature $R$ are given by $R_{\alpha\beta}=\frac{1}{2}%
\eta_{\alpha\beta}\square\phi+\frac{(N-2)}{2}(\partial_{\alpha}\partial
_{\beta}\phi+\frac{1}{2}\eta_{\alpha\beta}\partial_{\sigma}\phi\partial
^{\sigma}\phi-\frac{1}{2}\partial_{\alpha}\phi\partial_{\beta}\phi)$ and
$R=e^{-\phi}[(N-1)\square\phi+\frac{1}{4}(N-1)(N-2)\partial_{\sigma}%
\phi\partial^{\sigma}\phi]$.}
\begin{equation}
\partial_{\mu}\partial_{\nu}\phi-\eta_{\mu\nu}\,\square\phi-\frac{1}%
{2}\partial_{\mu}\phi\partial_{\nu}\phi-\frac{1}{4}\eta_{\mu\nu}%
\partial_{\alpha}\phi\partial^{\alpha}\phi=-\kappa T_{\mu\nu},
\label{Einstein scalar}%
\end{equation}
where $\square$ denotes the d'Alembertian operator in Minkowski spacetime and
$\partial^{\alpha}\phi=\eta^{\alpha\beta}\partial_{\beta}\phi$ . At this point
let us note that (\ref{Einstein scalar}) may be regarded as a dynamical
equation for a certain scalar field $\phi$ \ defined in a flat spacetime
background. From the standpoint of the Weyl frame observers \ this could be a
perfectly possible interpretation. Note that, although in this frame the
Riemannian curvature has been removed away, no information has been lost with
regard to geodesic motion.

\ Let us return to the question of the weak field approximation. Again,
recalling that in this approximation $\phi$ is considered static and small,
hence neglecting quadratic terms in the derivatives of $\phi$, Eq.
(\ref{Einstein scalar}) reduces to
\[
\partial_{\mu}\partial_{\nu}\phi-\eta_{\mu\nu}\,\square\phi=-\kappa T_{\mu\nu
}.
\]
On the other hand, for a perfect fluid configuration (defined in Minkowski
spacetime) we have\ $T_{\mu\nu}=(\rho c^{2}+p)V_{\mu}V_{\nu}-p\eta_{\mu\nu}$,
where $\rho$, $p$ and $V^{\mu}$ denotes, respectively, the proper rest mass
density, pressure and velocity field of the fluid. In a non-relativistic
regime we also neglect $p$ with respect to $\rho$, which implies that
$T_{00}\simeq\rho c^{2}$. For $\mu=\nu=0$ the above equation gives
\[
\nabla^{2}\phi=-\kappa\rho c^{2}%
\]
From (\ref{Newtonian-potential}) and substituting $\kappa=\frac{8\pi G}{c^{4}%
}$ we obtain%
\[
\nabla^{2}U=-4\pi G\rho,
\]
which is Poisson's equation of Newtonian gravity. This seems to suggest that
the scalar field $\phi$ contains, in fact, all information regarding gravity
in both Riemannian and Weyl frames.

The Einstein equations written in the form of (\ref{Einstein scalar}) may be
useful to obtain exact solutions for conformally flat spacetimes once we know
the energy-momentum tensor corresponding to a given matter configuration.
Although it is tempting to regard them as dynamical equations for a scalar
field $\phi$ in Minkowski spacetime there is a strong objection to such an
interpretation: those equations cannot be derived from an action principle.
Clearly, a scalar field must obey a scalar equation, while
(\ref{Einstein scalar}) are tensor equations. There is, however, a scalar
equation naturally associated with $\phi$, which can be obtained by
contracting (\ref{Einstein scalar}) with the Minkowski metric $\eta_{\mu\nu}$.
This leads to
\begin{equation}
\square\phi+\frac{1}{2}\phi^{,\mu}\phi_{,\mu}=\frac{\kappa}{3}T,
\label{scalar gravity}%
\end{equation}
where $T=\eta^{\mu\nu}T_{\mu\nu}$ denotes the trace of $T_{\mu\nu}$ with
respect to Minkowski metric $\eta_{\mu\nu}$. Of course, the above equation is
equivalent to the scalar equation $R=\kappa T$, obtained by taking the trace
of (\ref{Einstein equation}) with respect to $g_{\mu\nu}=e^{\phi}\eta_{\mu\nu
}$. It is interesting that this equation can also be derived from an action
principle, as we will show next.

\section{A scalar equation for the Weyl scalar field}

Let us leave general relativity for a while and speculate how one would
formulate a strict scalar field theory of gravity in which the Weyl scalar
field, a purely geometrical entity that defines the affine connection, would
play the role of the gravitational field in a Minkowski background. We have
seen in \ the previous section that, as long as we restrict ourselves to
conformally flat spacetimes, gravity may effectively be described by a scalar
field in Minkowski spacetime. An interesting approach to this question would
be to start with the formulation of general relativity in terms of a
variational principle. Thus, let us consider the Einstein-Hilbert action of
the gravitational field in the presence of matter%
\begin{equation}
S=\int_{\Omega}\sqrt{-g}\left(  R+\kappa L_{m}\right)  d^{4}x\text{ ,}
\label{action}%
\end{equation}
where $R$ is the scalar curvature, $L_{m}$ denotes the Lagrangian density of
matter, $\kappa$ is the Einstein constant and $\Omega$ is a regular domain in
$M$. The duality between the Riemann frame $(M,e^{\phi}\eta,0)$ and the Weyl
frame $(M,\eta,-d\phi)$ seems to suggest that in the variation of the
functional (\ref{action}) we should consider only variations $\delta g_{\mu
\nu}$ restricted to the class of conformally flat spacetimes, that is,
variations of the form \
\begin{equation}
\delta g_{\mu\nu}=\ \delta(e^{\phi}\eta_{\mu\nu})=e^{\phi}\eta_{\mu\nu}%
\delta\phi. \label{variation}%
\end{equation}
It is then not difficult to verify that the variation $\delta S$ in the action
induced by (\ref{variation}) will be given by \
\begin{equation}
\delta S=-\int_{\Omega}e^{\phi}\left(  R_{\mu\nu}-\frac{1}{2}g_{\mu\nu
}R+\kappa T_{\mu\nu}\right)  \eta^{\mu\nu}\delta\phi d^{4}x\text{ ,}
\label{variation of action}%
\end{equation}
where we have taken into account that $g^{\mu\nu}=e^{-\phi}\eta^{\mu\nu}$,
$\delta g^{\mu\nu}=-e^{-\phi}\eta^{\mu\nu}\delta\phi$, $\sqrt{-g}=e^{2\phi}$
and, as usual, the energy-momentum tensor $T_{\mu\nu}$ is defined by
$\delta\int_{\Omega}\sqrt{-g}L_{m}d^{4}x=$ $\int_{\Omega}\sqrt{-g}T_{\mu\nu
}\delta g^{\mu\nu}d^{4}x$. Since $\delta\phi$ is arbitrary, the condition
$\delta S=0$ implies
\begin{equation}
R=\kappa\mathcal{T}, \label{Fokker}%
\end{equation}
where $\mathcal{T}=g^{\mu\nu}T_{\mu\nu}=e^{-\phi}\eta^{\mu\nu}T_{\mu\nu}$
denotes the trace of the energy-momentum tensor with respect to the metric
$g_{\mu\nu}$. We see, then, that we can derive (\ref{Fokker}) by just taking
the variation of (\ref{action}) with respect to $\phi$ in the restricted class
of conformally spacetime metrics. On the other hand, if we express the
curvature scalar $R$ in terms of $\phi$ we get
\begin{equation}
R=3e^{-\phi}(\square\phi+\frac{1}{2}\phi^{,\mu}\phi_{,\mu}), \label{scalar}%
\end{equation}
where we are now using the notation $\phi_{,\mu}=\partial_{\mu}\phi$ and
$\phi^{,\mu}=\eta^{\mu\nu}\phi_{,\nu}$.\ Substituting (\ref{scalar}) into
(\ref{Fokker}) yields again
\begin{equation}
\square\phi+\frac{1}{2}\phi^{,\mu}\phi_{,\mu}=\frac{\kappa}{3}T\text{ ,}
\label{scalar gravity/}%
\end{equation}
where, as before, $T=\eta^{\mu\nu}T_{\mu\nu}$ denotes the trace of $T_{\mu\nu
}$ with respect to Minkowski metric $\eta_{\mu\nu}$. We now can look at this
equation from the point of view of the Weyl frame $(M,\eta,-d\phi)$ and regard
it as a dynamical field equation for the scalar field $\phi.$ Finally, we note
that if we define the new scalar field variable $\psi=e^{\frac{\phi}{2}}$ we
can get rid of the quadratic term $\frac{1}{2}\phi^{,\mu}\phi_{,\mu}$, \ and
hence Eq.(\ref{scalar gravity})\ may be put in the simpler form
\begin{equation}
\frac{\square\psi}{\psi}=\frac{\kappa}{6}T\text{ .} \label{scalar 2}%
\end{equation}

At this point, we think it is worth mentioning that the scalar equation
(\ref{scalar gravity}) may be derived from varying the action
\begin{equation}
S=\int d^{4}xe^{\phi}[\partial_{\mu}\phi\partial^{\mu}\phi+\frac{2}{3}\kappa
T] \label{scalar action}%
\end{equation}
with respect to the scalar field $\phi$. It is also interesting to note that
scalar gravity does not couple with a purely radiating electromagnetic field
since in this case $T=0$, which is consistent with the fact that in the
Riemann frame spacetime is conformally flat, and so null geodesics
representing light rays consist of straight lines.

We would like to conclude this section with two comments. First, let us note
that Eq. (\ref{scalar gravity}) is a\ direct consequence of the Einstein's
field equations%
\begin{equation}
R_{\mu\nu}-\frac{1}{2}g_{\mu\nu}R=-\kappa T_{\mu\nu}, \label{Einstein}%
\end{equation}
which for conformally flat spacetimes in the Riemann frame takes the form of
Eq. (\ref{Einstein scalar}).Therefore, any solution of the Einstein equations
is also a solution of (\ref{scalar gravity}) since this equation comes from
(\ref{Fokker}). Nevertheless, the converse is not true, and this implies that
the class of solutions of (\ref{scalar gravity}) is larger than the class of
solutions of (\ref{Einstein scalar}). However, because (\ref{scalar gravity})
is much easier to solve, it may sometimes be helpful in getting solutions of
(\ref{Einstein scalar}).

Finally, let us note that if we include the cosmological constant $\Lambda$ in
the Einstein equations (\ref{Einstein}), then (\ref{Einstein scalar}) and
(\ref{scalar gravity}) will become, respectively,
\[
\partial_{\mu}\partial_{\nu}\phi-\eta_{\mu\nu}\,\square\phi-\frac{1}%
{2}\partial_{\mu}\phi\partial_{\nu}\phi-\frac{1}{4}\eta_{\mu\nu}%
\partial_{\alpha}\phi\partial^{\alpha}\phi-\Lambda\eta_{\mu\nu}e^{\phi
}=-\kappa T_{\mu\nu}\text{ ,}%
\]%
\begin{equation}
\square\phi+\frac{1}{2}\phi^{,\mu}\phi_{,\mu}+\frac{4}{3}\Lambda e^{\phi
}=\frac{\kappa}{3}T.\label{cosmological constant}%
\end{equation}
It is not difficult to see that in terms of $\psi=e^{\frac{\phi}{2}}$
(\ref{cosmological constant}) reads
\[
\square\psi+\frac{2}{3}\Lambda\psi^{3}=\frac{\kappa}{6}\psi T.
\]
Incidentally, we note that in vacuum ($T=0)$ the above equation is a
non-linear Klein-Gordon equation , which has exact solutions in the form of a
travelling wave \cite{Polyanin}.

\section{Weyl frames and scalar theories of gravity}

As is well known, scalar theories of gravity first appeared with the work G.
Nordstr\"{o}m \cite{Nordstrom}, in his attempts to formulate a special
relativistic theory of gravitation \cite{Murdoch}. In the second of these
attempts, he postulates the following equation for the gravitational field:
\begin{equation}
\square\Phi=-4\pi G\Phi\eta^{\alpha\beta}T_{\alpha\beta}\text{ ,}
\label{Nordstrom}%
\end{equation}
where $T_{\alpha\beta}$ is the energy-momentum tensor of the matter content
and spacetime is flat \cite{Wheeler}. Norsdstr\"{o}m also assumed that the
motion of test particles would obey the equation
\begin{equation}
\dot{V}_{\alpha}=-\partial_{\alpha}\Phi-\dot{\Phi}V_{\alpha}\text{ ,}
\label{Motion}%
\end{equation}
where dot denotes derivative with respect to proper time in Minkwoski space.
This theory, as was shown by Einstein and Fokker \cite{Einstein-Fokker}, may
be formulated in terms of a metric theory of gravity whose field equation is
\[
R=-24\pi G\mathcal{T}\text{ ,}%
\]
with $\mathcal{T}$ $\ $as defined in the previous section. There is a
supplementary condition, namely that the Weyl tensor constructed from
$g_{\alpha\beta}$ vanishes or, equivalently, that spacetime is conformal. In
the Einstein-Fokker approach, the equation of motion (\ref{Motion}) is
replaced by a geodesic equation with respect to the metric $g_{\alpha\beta
}=\Phi^{2}\eta_{\alpha\beta}$, which also defines proper time in this curved
space. It is interesting to note that both approaches may be\ formally
considered as leading to the same theory formulated in different frames:
Einstein's in the Riemann frame $(M,\Phi^{2}\eta,0)$, and Nordstr\"{o}m's in
the Weyl frame $(M,\eta,-\frac{2}{\Phi}d\Phi)$. Note that by putting
$\Phi=e^{\frac{\phi}{2}}=\psi$ we see that Eq. (\ref{Nordstrom}) is equivalent
to (\ref{scalar gravity}) or to (\ref{scalar 2}).

\section{Weyl frames and two-dimensional gravity}

It is a well known result of differential geomety that in two dimensions all
spaces are conformally flat. This fact makes the duality between the Riemann
and Weyl frames in two dimensions completely general. In other words,
geometrical phenomena taking place in a curved two-dimensional space may be
described in a flat space endowed with a Weyl connection by simply changing
frames. This feature of two-dimensional geometry may perhaps be explored in
the context of two-dimensional gravity models. In this Section, we would like
to glimpse gravity in this dimensionality in a Weyl frame.

Lower-dimensional theories of gravity, mainly in connection with the
quantization of the gravitational field program, have atracted the attention
of many physicist during the last forty years \cite{Brown}. One of the most
popular versions of two-dimensional gravity, which reduces to Newtonian
gravity in two-dimensional in non-relativistic and weak field regime,
postulates the field equation \cite{Mann}
\begin{equation}
R+\Lambda=8\pi G\mathcal{T} \label{Jackiw}%
\end{equation}
Because in two dimensions the Riemann tensor is completely determined by the
curvature scalar $R$, the above equation seems to be the natural analogue of
the Einstein\ equations with the cosmological constant $\Lambda$. Let us note
that in this theory the conservation laws $T_{\;\;\;\;;\nu}^{\mu\nu}=0$ cannot
be deduced from the field equation (\ref{Jackiw}) and has to be separately postulated.

For a conformally flat metric $g=e^{\phi}\eta$\ the curvature scalar $R$ is
given, in two dimensions, by $R=e^{-\phi}\square\phi$. We thus can write Eq.
(\ref{Jackiw}) as
\begin{equation}
\square\phi+e^{\phi}\Lambda=8\pi GT \label{twodimensional}%
\end{equation}
where, as in (\ref{scalar gravity}) and (\ref{cosmological constant}),
$T=\eta^{\mu\nu}T_{\mu\nu}$ denotes the trace of $T_{\mu\nu}$ with respect to
Minkowski metric $\eta_{\mu\nu}$. As before, in the Weyl frame
Eq.(\ref{twodimensional}) may be interpreted as a dynamical equation for the
Weyl scalar field $\phi$, which again plays the role of the gravitational
field. Incidentally, in the absence of the cosmological constant
Eq.(\ref{twodimensional}) becomes
\[
\square\phi=8\pi GT
\]
which is a wave equation with source term.

\section{Weyl frames and quantum gravity}

As a mathematical tool conformal transformations have been widely used in
general relativity, particularly in the theory of asymptotic flatness
\cite{Wald} . They also have been employed in connection with scalar-tensor
theories of gravity. In fact there has been a long debate on whether different
frames related by conformal transformations have equivalent physical meaning
\cite{Faraoni}. In \ different conformal frames, the description of physical
phenomena may look different, though they are related to each other by a
mathematical transformation. To our knowledge this debate has, apparently,
being restricted to the context of classical physics. In this section we would
like to very briefly discuss some ideas on the subject of quantization of the
gravitational field in connection with the notion of Weyl frames. Surely we
are not considering the problem of quantum gravity in its generality as we are
restricted to the class of conformally flat spacetimes.

Quantum gravity is widely recognized as one of the most difficult and
challenging problems of theoretical physics. There is currently a vast body of
knowledge which includes several approaches to this area of research, but as
far as we know none of them has been entirely succesful to date. Among the
most popular of these are string theory \cite{String} and loop quantum gravity
\cite{Rovelli}. There is, however, a feeling among theorists, that a final
theory of quantum gravity, if indeed there is one, is likely to emerge
gradually and will ultimately be a combination of different theoretical
frameworks. In this spirit let us indulge ourselves \ for a while in raising
some questions concerning the possibility of using of the concept Weyl frame
as a way of looking at the problem of quantization of the gravitational field
in some particular cases.

We begin by considering the class of all conformally flat spacetimes, i.e.
those for which $C_{\alpha\beta\mu\nu}=0$, where $C_{\alpha\beta\mu\nu}$
denote the components of Weyl's conformal tensor. This condition reduces the
number of independent metric components to only one, and also implies that
$g=e^{\phi}\eta$, where $\phi$ is a scalar field. Now, this kind of geometry,
whose number of degrees of freedom has been drastically reduced\ and fixed
\textit{a priori }is an example of what has been called a \textit{prior
geometry }\cite{Wheeler}\textit{. }For such spacetimes all information about
the gravitational field is encoded in the scalar field $\phi$, so it seems not
unreasonable to expect that any quantum aspect emerging in the process of
quantization of the gravitation field should somehow involve this field, even
though it may be objected that gravitons generated by $\phi$ would be spin-0
particles . Moreover, one would also expect that the correspondence between
the Riemann and Weyl frames, which holds for conformally flat spacetimes at
the classical level, would be preserved at the quantum level. If this is true,
then it would make sense to carry over the scheme of quantization from the
Riemann frame to the Weyl frame. Nevertheless, although in the Weyl frame the
spacetime geometry has no longer any degrees of freedom, the scalar field
$\phi$ is still a repository of physical information. It would then seem
plausible to treat $\phi$ as a physical field. We then are left with a
situation which is typical of the ones considered by quantum field theory in
flat spacetime. In fact this is not so unusual as in perturbative string
theory spacetime is also treated as an essentially classical background
\cite{String}, not to mention that Feymann used to hold the view that a
quantum theory of gravitation should lead to massless spin-2 quanta coupled to
matter in flat Minkowski spacetime \cite{Feynmann}. Let us suppose that we
succumb to the temptation of pursueing this analogy more seriously and proceed
to the quantization of the Weyl scalar field in Minkowski spacetime. Many
questions would arise at this point. For instance, to quantize the scalar
field $\phi$ we need to know what the dynamics of $\phi$. We have seen that
the dynamical equations of $\phi$ are given by (\ref{Einstein scalar}) of
which (\ref{scalar gravity}) is a consequence. What would happen, however, if
we provisionally regarded (\ref{scalar gravity}), which may be derived from
the action (\ref{scalar action}), as the fundamental equation to be quantized?
We leave this and other questions for future work.

\section{ Conclusion}

In this paper we have developed the idea that there is a connection between
two different geometrical descriptions of conformally flat spacetimes. With
the help of concepts borrowed from Weyl geometry we have shown that some
geometrical phenomena taking place in a Riemannian curved spacetime may be
described in Minkowski flat space, in such a way that the curvature\ of the
first is replaced, in the second, by a dynamical scalar field $\phi$. This
field has a geometrical character as it gives rise to a non-Riemannian affine
connection in the flat space. We are thus left with two different pictures.
Accordingly, we can reinterpret the essential facts of Robertson-Walker
cosmology in terms of a flat spacetime cosmology, in which the motion of
galaxies takes place in Minkowski spacetime and is determined by a scalar
field. A similar scenario was conceived some years ago, but in a different
context in which there is no scalar field but the particle masses may depend
on space and time \cite{Narlikar}. We have also discussed other concrete
situations in which the two mathematically equivalent descriptions seem to
lead to different physical pictures, and these are scalar gravity and
two-dimensional gravity. Finally, we slightly touch on the possibility of
investigating whether one could apply the mathematical formalism connecting
conformally flat spacetimes and scalar fields to the quantization of gravity
in Minkowski spacetime.

\section{Acknowledgement}

The authors would like to thank CNPq and CLAF for financial support.

\end{document}